\def\be{\begin{equation}}
\def\ee{\end{equation}}
\def\bea{\begin{eqnarray}}
\def\eea{\end{eqnarray}}

\documentclass[12pt]{article}
\usepackage{epsfig,latexsym}
\begin{document}
\thispagestyle{empty}
\vspace*{-0.5 cm}
\vspace*{-1.2in}
\begin{flushright}
\end{flushright}
\vspace*{0.7 in}
\begin{center}
{\large \bf Towards an accurate determination of the critical exponents 
with the Renormalization Group flow equations} \\
\vspace*{1cm} {\bf A. Bonanno} \\ \vspace*{.3cm} 
{\it Osservatorio Astrofisico} \\ 
{\it Via S.Sofia 78, I-95138, Catania} \\
{\it INFN, Sezione di Catania} \\
{\it Corso Italia 57, I-95129, Catania, Italy} \\

\vspace*{.3cm}
 and \\
\vspace*{0.3cm}
{\bf D. Zappal\`a} \\ \vspace*{.3cm} 
{\it INFN, Sezione di Catania} \\ 
{\it Dipartimento di Fisica, Universit\`a di Catania}\\
{\it Corso Italia 57, I-95129, Catania, Italy} \\

\vspace*{1 cm}
{\bf ABSTRACT} \\
\end{center}
The determination of the critical exponents  
by means of the Exact Renormalizion Group approach 
is still a topic of debate.
The general flow equation is by construction scheme 
independent, but
the use of the truncated derivative expansion generates a 
model dependence in
the determination of the universal quantities.
We derive new nonperturbative flow equations 
for the one-component, $Z_2$ symmetric scalar field
to the next-to-leading order of the derivative expansion
by means of a class of proper time regulators.
The critical exponents $\eta$, $\nu$ and $\omega$ 
for the Wilson-Fisher fixed point are computed by numerical 
integration of the flow equations, without resorting to 
polynomial truncations. 
We show that by reducing the width of the cut-off employed, 
the critical exponents become rapidly insensitive to the 
cut-off width and their  values  are in good
agreement  with the results of entirely different approaches.
\\
\vskip 0.5 cm
\noindent
Pacs 11.10.Hi , 11.10.Kk

\parskip 0.3cm
\vspace*{3cm}
\vfill\eject
\setcounter{page}{1}
\voffset -1in
\vskip2.0cm

\newcommand{\fa}{\phi^a}
\newcommand{\fb}{\phi^b}
\newcommand{\p}{\partial_{\mu}}
\newcommand{\dd}{\delta^{ab}}
\newcommand{\nn}{\nonumber}

{\it Universality} is the property which guarantees that,
at a critical point, the divergences of the correlation functions
of an entire  class of physical systems have the same 
power law behavior with the same exponents (critical 
exponents), regardless of their specific features and 
especially of their microscopic (ultraviolet (UV)) details. 
This notion is very general and it often appears in the study of 
critical phenomena and in quantum field theory. The prediction of 
those exponents has been recognized to be one of the most important 
achievements of the renormalization group approach pioneered by Wilson 
and others \cite{kad,wil}. A standard approach to this 
problem is the $\epsilon$-expansion \cite{wil} method, 
whose main feature is an expansion in $\epsilon=4-D$ (where $D$ 
indicates the number of dimensions) of the renormalized free energy. 
Unfortunately this method is not entirely satisfactory, 
since actual values for three-dimensional systems are obtained only 
after setting $\epsilon=1$ at the end of the calculation. 
On the other hand, ordinary perturbation theory in such nonperturbative 
framework requires quite involved loop calculations \cite{gzj}. 
Moreover since both $\epsilon$-expansion and perturbative expansion at 
fixed dimension yield divergent series, 
one has to rely on the Borel summability of these series 
in order to get meaningful results.

In recent years
the Exact Renormalization Group (ERG) approach has proven 
to be very useful in many situations where the usual perturbative methods
are not totally satisfactory \cite{berg,rep1}. The main
step in this approach is to solve the flow equation which 
is a complicate non-linear functional equation. For this reason  
one must then resort to sofisticate approximation methods
in order to extract physical information.
The derivative expansion allows a systematic expansion of 
the renormalized action in terms of local field operators. 
It has the advantage of recasting 
the functional flow equation into a set 
of coupled partial non-linear differential equations for the local potential
$U_k(\phi)$ and for the wavefunction renormalization function 
$Z_k(\phi)$ in the next-to-leading order approximation which improves 
the so called Local Potential Approximation (LPA) where only the 
equation for $U_k(\phi)$ is considered and $Z_k(\phi)=1$ is kept fixed.   
This topic has been the subject of an intense investigation and the reader is
referred to the excellent review in \cite{rep2} for a 
recent survey of results for the scalar theory.
Although, as far as the calculation of universal quantities is concerned,
the exact evolution equation is by construction scheme independent,
this no longer holds in specific truncations of the full evolution equation. 
In fact different realizations of the smooth cut-off ERG equations determine 
critical exponents which are different if, for instance,  an exponential 
cut-off \cite{wette} or a power law cut-off is used \cite{morris,morrispl}.

Recently, the ERG scheme dependence has been the subject of many papers.
In \cite{litio} the scheme dependence has been analyzed in order to optimize the 
convergence of approximate solutions of the ERG equations. 
Other authors have considered the effects of the scheme dependence in the convergence 
of polynomial truncations of the flow equation in the LPA \cite{polia,papp}. 
Actually it would be desirable to have an approximation framework 
which allows to control the scheme dependence to each order in  
the derivative expansion, without resorting to further polynomial 
truncations and which, at the same time, could provide  an estimate 
of the uncertainty on the determination of the universal quantities.

The first determination of the critical exponents $\nu,\omega$
from a numerical resolution of the flow equation has been performed in 
\cite{hasen} by using the LPA of the Wegner-Houghton (WH) equation for the 
wilsonian action,\cite{wegner}, which is a sharp cut-off version of the ERG equation.
However the WH equation, truncated to the next-to-leading order in the derivative 
expansion although  predicts the correct two loop value of the field anomalous 
dimension $\eta$ in $D=4$ \cite{noi}, fails to give a reasonable value of $\eta$ 
at the Wilson-Fisher fixed point in $D=3$ \cite{bbmz}.

To overcome the difficulties with the sharp cut-off formulation, several 
authors \cite{olesc,percacci,liaooc} have proposed a smooth cut-off 
realization of the ERG equations by means of the Schwinger proper time
regulator. This approach  has been used in many contexts like Yang-Mills 
theories \cite{liaoym} and Chiral Symmetry Breaking \cite{papp,pirnertutti}.
In \cite{papp} the effect of this regulator on specific polynomial truncations
of the LPA has been discussed. In \cite{bohr} the same cut-off function  
is used to include the effect of a field independent 
wavefunction renormalization 
function $Z_k$ in the $O(N)$ model. However the inclusion of the field 
dependence in $Z_k$ turns out to be essential for a consistent determination 
of the scaling properties of the system. 
A discussion on the determination of the wavefunction renormalization 
function within this framework has been presented even in \cite{liaooc},
but so far the  full flow equation for $Z_k(\phi)$ within this approach
has not been derived.

The aim of this paper is to derive the full coupled flow equations for 
$Z_k(\phi)$ and $U_k(\phi)$ by means of the general class of proper time 
regulators introduced in \cite{olesc,percacci,liaooc} and to numerically 
determine the anomalous dimension $\eta$ and the critical exponents 
$\nu$ and $\omega$, by solving these equations around the Wilson-Fisher 
fixed point in $D=3$.
We also show that, when diminishing the size of the cut-off width,
the critical exponents become stable, converging to fixed values.

Let us now review the basic assumptions in the derivation of our
evolution equation. 
The one-loop contribution to the effective action is given by the trace of 
${\rm log}(\delta^2 S/( \delta\phi\delta\phi))$, where $S$ is the 
classical action. 
By introducing the Schwinger proper time representation for the 
logarithm with the proper time variable $s$, we can express the one-loop 
contribution as 
\be
\label{eq:g1l}
S^{1-loop}_k = -{1\over 2} {\rm Tr} \; \int_0^\infty \;{ds\over s} f_k \; 
{\rm exp}\; \Big ({-s{\delta^2 S \over 
\delta\phi\delta\phi}} \Big )
\ee
where we have introduced the heat-kernel smooth regulator $f_k$.
The usual one-loop effective action is recovered by setting $f_k=1$ and
properly regulating the UV divergences in Eq. (\ref{eq:g1l}).

In the spirit of the derivation of the WH equation, the flow of the 
wilsonian action is obtained by taking $f_k$ as a sharp 
momentum cut-off, i.e. a Heaviside function which is equal to one  for momenta
greater than $k$ and vanishes for momenta smaller than $k$, 
and then by performing the integration of the modes of the action through 
infinitesimal steps $\delta k$ in the momentum space.
Therefore the action $S_{k-\delta k}$ comes from the integration of the modes
between $k-\delta k$ and $k$ in the action  $S_k$ defined at the scale $k$.
A smooth cut-off on the momenta in this framework is then obtained 
by choosing a non-singular $f_k$ instead of the Heaviside function
and the differential flow equation reads
\be
\label{eq:gamma}
k\;{\partial S_k \over \partial k }=
-{1\over 2} {\rm Tr} \; \int_0^\infty \;{ds\over s} \;
k {\partial f_k \over \partial k}\;
{\rm exp}\; \Big (-s{\delta^2 S_k \over 
\delta \phi\delta \phi} \Big )   
\ee

It must be remarked that the WH equation contains terms that are neglected 
in Eq. (\ref{eq:gamma}) but, as discussed in \cite{polo}, these terms do not
contribute to the next-to-leading truncation in the derivative expansion. 

About $f_k$ one can in general require the following properties (see 
\cite{percacci,liaooc}):
1) $f_k$ must depend on $k$ through the dimensionless quantity $s k^2$.
We shall consider, as in \cite{bohr} the variable $s Z_k k^2$,
where $Z_k$ is the wave function renormalization function, in order to 
reproduce the usual form of the propagator in the following equations.
2) $f_{k=0}=1$ so that  $S_{k=0}$ must be identified with  
the full effective action. This implies that 
in the limit $k\rightarrow 0$ the infrared cut-off is removed. 3) 
$f_k(s Z_k k^2)$ must tend to zero sufficiently rapidly  for large 
$s Z_k k^2$ to suppress the small momentum modes. 
In particular we choose, for any integer  $m\geq 1$
\be
\label{eq:effe}
f^{(m)}_k(s Z_k k^2)=e^{-s Z_k k^2}\sum^m_{i=0}{{(sZ_k k^2)^i}\over {i!}}
\ee 

We note that, as it stands, Eq. (\ref{eq:gamma})
for some values of $m$ and of the dimension $D$ is not UV convergent,
but in $D=3$ and for  $m\ge 1$, which is the case we are interested in,
no UV divergence appear in the differential flow equation.
 
An analysis of the polynomial truncations of the
LPA to Eq. (\ref{eq:gamma}) for $m=1,2,3$  has been
performed in \cite{papp}, where it is shown that all the parameters 
extracted from the flow equations, regarded as functions of the polynomial
truncation, converge more rapidly when $m$ grows. This is a first
indication that the RG equations provide more stable results  for
increasing values of $m$.

As anticipated above, we shall focus here on the first order truncation 
in the derivative expansion of the action $S_k$
which corresponds to the following parametrization
\be
\label{eq:formgam}
S_k=\int d^Dx \Big ( {{Z_k(\phi)}\over  {2}}\partial_\mu \phi
\partial^\mu\phi+V_k(\phi)\Big )
\ee
and therefore Eq. (\ref{eq:gamma}) is reduced to two coupled 
partial differential equations for $V_k(\phi)$ and $Z_k(\phi)$. No
further polynomial expansion or truncation in powers of the field $\phi$
is made on  $V_k(\phi)$ and
$Z_k(\phi)$.

In order to project the flow equation on the local potential term and on the
kinetic term we first set (see also \cite{fraser,polo,noi}) 
$\phi=\phi_0+\varphi(x)$ where $\phi_0$ is a constant field configuration 
and $\varphi(x)$ is a small fluctuation field. Next, we expand $V_k(\phi)$
and $Z_k(\phi)$ in both sides of Eq. (\ref{eq:gamma}) and retain only terms
at most quadratic in $\partial\varphi$. Then we define
$\delta^2 S_k/(\delta\phi\delta\phi)\equiv A=A_0+\delta A$ 
where $A_0$ is only a
function of the constant field $\phi_0$ and  $\delta A$ contains all the
linear and quadratic contributions in  $\partial \varphi$.  Thus the trace 
becomes
\bea\label{eq:3}
&&-{1\over 2} {\rm Tr} \; \int_0^\infty \;{ds\over s} \;
k {\partial f_k \over \partial k}\; e^{-sA}=  \nonumber\\
&&\int d^D x 
\int_0^\infty {ds} \; Z_k k^2 {(sZ_k k^2)^m\over m!} e^{-sZ_k k^2}\langle x|
e^{-s(A_0+\delta A)}| x \rangle =\nonumber\\
&&\int d^D x 
\int_0^\infty {ds} \; Z_k k^2 {(sZ_k k^2)^m\over m!} e^{-sZ_k k^2}\langle x|
e^{-sA_0}(1-s\delta A\nonumber\\
&&+{s^2\over 2!}\{[\delta A,A_0]+\delta A^2]\}+... )| x \rangle
\eea
where in order to disentangle the trace in Eq. (\ref{eq:3})
we used the Baker-Campbell-Hausdorff formula, 
and the dots stand for the higher order terms
in the $s$ expansion of the exponential. Actually, in order to 
collect all the contributions to the wavefunction renormalization
function we had to retain terms up to the $O(s^4)$ order in this expansion.
The evaluation of the trace, as explained in \cite{fraser,polo,noi},
is then  performed by inserting the identity $\int d p^D | p \rangle
\langle p |$ in the trace in Eq. (\ref{eq:3})
and by making use of the commutation relation 
$[p_\mu, f(x)]=i\partial_\mu f(x)$. Finally, by collecting on both sides of 
Eq. (\ref{eq:3}) the coordinate independent terms and the coefficients of the
kinetic term $\partial\varphi \partial\varphi$, we get respectively 
the flow equation for $V_k$ and for $Z_k$.
Remarkably, all the integrals in the proper time $s$ can be 
performed analytically and therefore we end up not with 
integrodifferential flow equations, but simply with differential equations.

It is convenient to express the flow equations in terms of new dimensionless
quantities $t,~\Phi,~V(t,\Phi),~Z(t,\Phi)$, defined by the following
relations: $t={\rm ln}(\Lambda/k)$,~
$\phi = k^{(1+\eta)/2}\sqrt{\alpha}\Phi$,~
$V_k(k,\phi)=\alpha k^{3}V(t,\Phi)$ and $Z_k(k,\phi)=k^{-\eta}Z(t,\Phi)$
where $\eta$ is the anomalous 
dimension of $\phi$, $\Lambda$ is a UV reference momentum scale,
and the constant $\alpha$ is defined in terms of the Gamma functions:
$\alpha=(\Gamma(m-1/2))/(8\pi\sqrt{\pi}m!)$). The flow equations then read
(the $\prime$ indicates derivative w.r.t. the field $\Phi$) 
\be
\label{eq:uflow}
{{\partial V}\over{\partial t}} = 3V - {{1+\eta}\over 2}\Phi V' -
\Biggl({{Z}\over{Z+V''}}\Biggr)^{m-1/2}
\ee
\bea
\label{eq:zflow}
{{\partial Z}\over{\partial t}} =-\eta Z &-& {{1+\eta}\over 2}\Phi Z' +
\Biggl({{Z}\over{Z+V''}}\Biggr)^{m-1/2}
\Biggl({{(m-1/2)}\over{Z+V''}} \biggl(Z''-{{49 (Z')^2}\over{24Z}}\biggr)-
\nonumber\\
&&{{7(m^2-1/4)}\over{6 (Z+V'')^2}}Z'V'''+
{{(m^2-1/4)(m+3/2)}\over{6(Z+V'')^3}}Z(V''')^2\Biggr )
\eea

In order to determine the critical exponents we need to find the  
fixed points of Eqs. (\ref{eq:uflow},\ref{eq:zflow}) and then
to study the linearized version of the equations around the fixed point 
solutions. The latter are the $t$-independent solutions of  
Eqs. (\ref{eq:uflow},\ref{eq:zflow}). They are indicated
as $V^*(\Phi)$ and $Z^*(\Phi)$. Once  $V^*(\Phi)$ and $Z^*(\Phi)$ are
determined, one can linearize Eqs. (\ref{eq:uflow},\ref{eq:zflow}), 
by introducing
small $t$-dependent perturbations around $V^*(\Phi)$ and $Z^*(\Phi)$:
\bea
\label{eq:pert}
V(t,\Phi)=V^*(\Phi)+\delta V(t,\Phi)=
V^*(\Phi)+e^{\lambda t}v(\Phi)\nonumber \\
Z(t,\Phi)=Z^*(\Phi)+\delta Z(t,\Phi)=
Z^*(\Phi)+e^{\lambda t}z(\Phi)
\eea 
and by retaining in  Eqs. (\ref{eq:uflow},\ref{eq:zflow}) only linear terms in
$v(\Phi),z(\Phi)$.

It is worthwhile to notice that 
Eqs. (\ref{eq:uflow},\ref{eq:zflow}) are invariant
under the transformations $\Phi\to \epsilon \Phi$, 
$Z\to \epsilon^{-2}Z$.
This is the consequence of a reparametrization 
invariance of Eq. (\ref{eq:gamma}) where
$\phi$ and $Z_k$ transform as $\Phi$ and $Z$ 
respectively and the proper time is redefined
as $s\to \epsilon^2 s$, and all the other quantities are not modified.

Before going on  we consider in more detail 
the behavior of the regulator $f_k$ when $m$ grows.
As a simple check one can go back to Eq. (\ref{eq:3}) and neglect the
$Z_k$ effects. In this case $\delta A=0$ and no expansion of the exponential
is needed and only the term ${\rm exp}(-sA_0)$ is left. By integrating
the variable $s$ we get
\bea
\label{eq:4}
-{1\over 2} {\rm Tr} \; \int_0^\infty \;{ds\over s} \;
k{\partial f_k \over \partial k}\; &e^{-sA_0}&=
\int d^3 p
\int_0^\infty {ds} \; k^2 {(s k^2)^m\over m!} e^{-s(k^2+A_0)}=
\nonumber\\
&&4\pi~\int_0^\infty dp p^2\Biggl ({{k^2}\over{k^2+p^2+{V_k}''}}\Biggr )^{m+1}
\eea

Clearly the $p$ integration leads back to Eq. (\ref{eq:uflow}) with $Z=1$,
but from the integrand in the r.h.s.  of Eq. (\ref{eq:4}) it is possible to 
learn which momenta $p$ 
are relevant in the integration. In Fig. 1 the rescaled integrand
$F(p)= m p^2 (k^2/(k^2+p^2+{V_k}'') )^{m+1}$
for some values of $m$ and for fixed $k$ is plotted. The maximum of $F(p)$
is at ${\overline p}^2=(k^2+{V_k}'')/m $ 
and $F(\overline p) =(k^2+{V_k}'')^{-m}(m k^2/(m+1))^{m+1}$.
So, when $m$ grows, the cut-off $f_k $ has the effect of selecting 
smaller and smaller momentum shells,
centered at $\overline p$ where the integrand behaves as a power of the
propagator. Then, the large $m$ limit is a kind of sharp cut-off
limit for a modified propagator, which however cannot be 
directly related to the
sharp cut-off WH equation. Indeed the logarithmic behavior of the latter is
clearly different from the one of Eq. (\ref{eq:uflow}) where the term
$(Z/(Z+V''))^{m-1/2}$ appears instead.
At the same time no straightforward limit
$m\to \infty$ can be taken in Eqs. (\ref{eq:uflow},\ref{eq:zflow})
in order to determine the
analytic asymptotic behavior of the critical exponents.
This can be seen in the LPA from Eq. (\ref{eq:uflow}) with $Z=1$.
In this case, when  $m\to \infty$, the sign of $V''$ determines whether the 
last term in (\ref{eq:uflow}) diverges or vanishes (or it is one for $V''=0$).
Therefore no uniform convergence of the solution is found in the limit 
$m\to \infty$.
Unfortunately, a deeper understanding of the relation between 
this heat-kernel 
regulator and the other formulations of the ERG equations
is still missing and we have to limit ourselves to compare the numerical 
results of the different methods.

We are interested in determining the anomalous dimension $\eta$ 
at the only non-gaussian fixed point solution, the Wilson-Fisher fixed point, 
and the exponents $\nu$ and $\omega$ which 
are related to the eigenvalues $\lambda$
of the linearized equations, being defined  as the
inverse of the only positive eigenvalue and as the opposite of the  
less negative eigenvalue respectively.

The first step concerns the determination of non-trivial solutions 
$V^*(\Phi)$ and $Z^*(\Phi)$ and we closely follow
the procedure outlined in \cite{morrispl} (see also \cite{bbmz}). 
We use the shooting method in order to numerically solve the fixed point 
equations taking, as boundary conditions, the asymptotic expression of the 
potential and wave function renormalization function at large 
$\Phi$ together with the symmetry property of these functions
at the origin: ${Z^*}'(0)={V^*}'(0)=0$. The normalization at the origin
$Z^*(0)$ must be fixed to some definite value and, as a test of the
reparametrization invariance, we checked that the numerical results found
for the critical exponents do not depend on the particular value 
of this normalization. The procedure to determine the
asymptotic expressions from the equations is explained in detail in
\cite{morrispl} and therefore we do not repeat it here.
In practice, as it happens for the sharp cut-off equations, 
\cite {bbmz}, we find that the numerical resolution of the problem is 
simpler if the equation for the potential is replaced by the
corresponding equation for its field derivative ${V^*}'(\Phi)$.
For each value of $m$ we find only one non-trivial fixed point as it was 
expected and  the values of the anomalous dimension for each $m$ 
are displayed in Table 1.

\begin{table} [h] \centering{
\begin{tabular}{|c|c||c|c|}  \hline
$m$ &\multicolumn{1}{c||}{$\eta$} & $m$ &\multicolumn{1}{c|}{$\eta$} \\
\hline 
$ 1$&$    0.0653$&$11$&$    0.0365$ \\
$ 2$&$    0.0507$&$12$&$    0.0362$ \\
$ 3$&$    0.0452$&$13$&$    0.0360$ \\
$ 4$&$    0.0423$&$14$&$    0.0358$ \\
$ 5$&$    0.0405$&$15$&$    0.0356$ \\
$ 6$&$    0.0393$&$16$&$    0.0354$ \\
$ 7$&$    0.0385$&$17$&$    0.0353$ \\
$ 8$&$    0.0378$&$20$&$    0.0350$ \\
$ 9$&$    0.0373$&$30$&$    0.0343$ \\
$10$&$    0.0369$&$40$&$    0.0340$ \\
\hline
\end{tabular}
\caption{\em
The anomalous dimension $\eta$ at the Wilson-Fisher fixed point
determined for various values of $m$.
}
\label{uno}
}
\end{table}

Having determined the fixed point, we again use the shooting procedure to 
evaluate $\nu$ and $\omega$ from the linearized equations as explained in 
\cite{morrispl}. The results are reported in Table 2 as the 
$O(\partial^2)$ approximation estimates.
In order to test the improvement due to the inclusion of the wave function 
renormalization function, we also show in Table 2 the values of 
$\nu,\omega$ obtained within the LPA
(~$O(\partial^0)$~), i.e. from the linearized potential equation
with $\eta=0$ and $Z=1$ fixed.

We note that the numerical procedure employed to solve the equations becomes 
more problematical and converges less rapidly for larger values of $m$. 
Moreover, since  the fixed point equations are more stable than the 
linearized ones we can push $m$ up to $m=40$ in the former case but only 
up to $m=9$ in the  latter case. 

\begin{table} [h] \centering{
\begin{tabular}{|c||c|c||c|c||c|c|c|}  \hline
$m$ &\multicolumn{2}{c||}{$\nu$}&\multicolumn{2}{c||}{$\omega$} \\ 
\hline 
&$O(\partial^0)$ & $O(\partial^2)$ & $O(\partial^0)$ &$O(\partial^2)$ \\ 
\hline
$ 1$&$  0.6604$&$   0.6348$&$ 0.629$&$  0.847 $\\
$ 2$&$  0.6439$&$   0.6331$&$ 0.674$&$  0.762 $\\  
$ 3$&$  0.6381$&$   0.6311$&$ 0.698$&$  0.738 $\\  
$ 4$&$  0.6351$&$   0.6299$&$ 0.711$&$  0.727 $\\ 
$ 5$&$  0.6333$&$   0.6290$&$ 0.720$&$  0.721 $\\ 
$ 6$&$  0.6322$&$   0.6286$&$ 0.727$&$  0.717 $\\  
$ 7$&$  0.6313$&$   0.6279$&$ 0.731$&$  0.713 $\\
$ 8$&$  0.6306$&$   0.6277$&$ 0.735$&$  0.711 $\\
$ 9$&$  0.6301$&$   0.6276$&$ 0.738$&$  0.710 $\\ 
\hline
\end{tabular}
\caption{\em
Values of the exponents $\nu,\omega$ for various $m$.
The $O(\partial^0)$ (LPA) values are obtained from the potential
equation only with $Z=1,~\eta=0$ fixed.
}
\label{due}
}
\end{table}

The first thing that should be noted is that the change  
of $\eta,\nu,\omega$ diminishes when $m$ grows
and the value of these parameters becomes almost constant :
$\nu$ and $\omega$ ($O(\partial^2)$) already at $m=9$ are  
almost stable up to the third significant digit  and, according 
to the trend shown up to $m=40$, a rapid convergence of
$\eta$ is observed. The dependence on the parameter $m$ is practically vanishing.
Since the analytical approach to the $m\to \infty$ limit of Eqs. 
(\ref{eq:uflow},\ref{eq:zflow}) is troublesome, we have to rely on the 
numerical results shown in Tables 1,2 in order to understand the behavior 
of $\eta,\nu,\omega$ for large values of $m$.
As an additional check we have considered a fit to the numbers in 
Table 1 with various analytical trial forms of the function $\eta(m)$,
containing a few free parameters. In these checks the $\chi$-square
analysis clearly indicates a finite asymptotic value of $\eta(m\to\infty)$ around
$\eta=0.0329$. 

As far as the derivative expansion improvement is concerned,
$\nu$ and $\omega$ behave differently in the two approximations
here considered: the former is always decreasing with $m$;
the latter has a different trend in the two approximations
and the LPA gives a better estimate than  the $O(\partial^2)$ approximation.
However it must be remarked that both $\nu,\omega$ become more stable 
with  $m$ when both  equations for $v$ and $z$ are considered, rather than in
the LPA.
It is conceivable to expect that, when going to higher orders in
the derivative expansion, the critical
exponents converge to a value that sits between
the $O(\partial^0)$ and the $O(\partial^2)$ estimates.

In Table 3 we report some estimates of the critical 
exponents obtained by various formulations of the ERG equations 
and by completely different methods (see captions for references).
Table 3 collects only few relevant results and it is by far an incomplete 
list. Much more exhaustive lists can be found in \cite{gzj,rep1,rep2,
hasenbush}.  
A comparison with the results in Tables 1 and 2 indicates that the 
particular cut-off here employed provides a little improvement in the
agreement with the non-ERG approaches, 
on the anomalous dimension determination.
Concerning $\nu$ and $\omega$ our determinations 
are within the average, but slightly smaller
(although it should be noticed the small experimental value of $\nu$).

\begin{table} [h] \centering{
\begin{tabular}
{|c|c|c|c|}
\hline
     & $\eta      $& $\nu$        &$\omega$     \\
\hline                                           
 (a) & $0.0467    $& $0.6307    $ &             \\
 (b) & $0.0539    $& $0.6181    $ &$0.897    $  \\
 (c) & $0.040(7)  $& $0.626(9)  $ &$0.85(7)  $  \\
 (d) & $0.042     $& $0.622     $ &$0.754    $  \\
\hline\hline
 (e) & $0.0335(25)$& $0.6304(13)$ &$0.799(11)$  \\
 (f) & $0.0360(50)$& $0.6290(25)$ &$0.814(18)$  \\
 (g) &             & $0.6300(15)$ &$0.825(50)$  \\
 (h) &             & $0.6310(5) $ &             \\
 (i) & $0.026(3)  $& $0.624(2)  $ &$0.82(3)  $  \\
 (l) & $0.0374(14)$& $0.6294(9) $ &$0.87(10) $  \\
\hline
\hline
 (p) &             & $0.625(6)  $ &$0.80(5)$    \\
\hline
\end{tabular} 
\caption{\em
Values of the exponents $\eta,\nu,\omega$, calculated with various
methods.\protect\\
(a) RG effective average action approach with exponential smooth cut-off, 
to  $O(\partial^2)$ order \cite{rep1}. 
\protect\\
(b) ERGE with powerlike smooth cut-off to $O(\partial^2)$ order 
\cite{morrispl,morristur}.
\protect\\
(c) ERGE expansion into a truncated set of ordinary differential equation
\cite{golner}, quoted from  \cite{rep2}.
\protect\\
(d) Polchinski version of the ERGE \cite{comellas}.
\protect\\
(e) Seven-loop perturbation series in $D=3$ \cite{gzj}.
\protect\\ 
(f) Five-loop $\epsilon$-expansion \cite{gzj}.
\protect\\ 
(g) high temperature series \cite{nickelrehr}, quoted from \cite{gzj}.
\protect\\
(h) high temperature series \cite{buteracomi}, quoted from \cite{gzj}.
\protect\\
(i) Monte Carlo simulation \cite{baillie}.
\protect\\
(l) Monte Carlo simulation \cite{ballesteros}, quoted from \cite{gzj}.
\protect\\ 
(p) experimental data from the liquid-vapour transition.
quoted from \cite{zjlibro}.
\protect\\
}
\label{tre}
}
\end{table}

Therefore, the convergence of the numerical results in Tables 1 and 2 to
such good values in the limit of large $m$, provides an indication that 
the heat-kernel cut-off, in that limit, becomes a sensible regulator
of both ultraviolet and infrared modes. Moreover it is particularly effective 
in reducing  the weight of  all the irrelevant operators neglected in our 
derivative expansion truncation, which, on the other hand, 
as it is argued in \cite{polia},
become important for the same truncation of the WH equation.

In conclusion, we have presented a particular version of 
the RG equations for a single field scalar theory with a heat-kernel 
cut-off,  introduced in the Schwinger proper time formalism. 
This cut-off is parametrized by an index that regulates the size of the 
momentum shell integrated in the blocking procedure.
We considered a derivative expansion of the flow 
equation truncated to the $O(\partial^2)$ order, 
obtaining a set of two coupled equations which are 
reparametrization invariant.
The corresponding determination of $\eta,\nu,\omega$ at the 
Wilson-Fisher fixed point in three dimensions is substantially encouraging 
due to the good agreement with the other  estimates of these parameters.

\vspace {1cm}

We would like to acknowledge Martin Reuter for very enlightening
discussions and encouragements.

\vfill\eject
\begin{figure}
\psfig{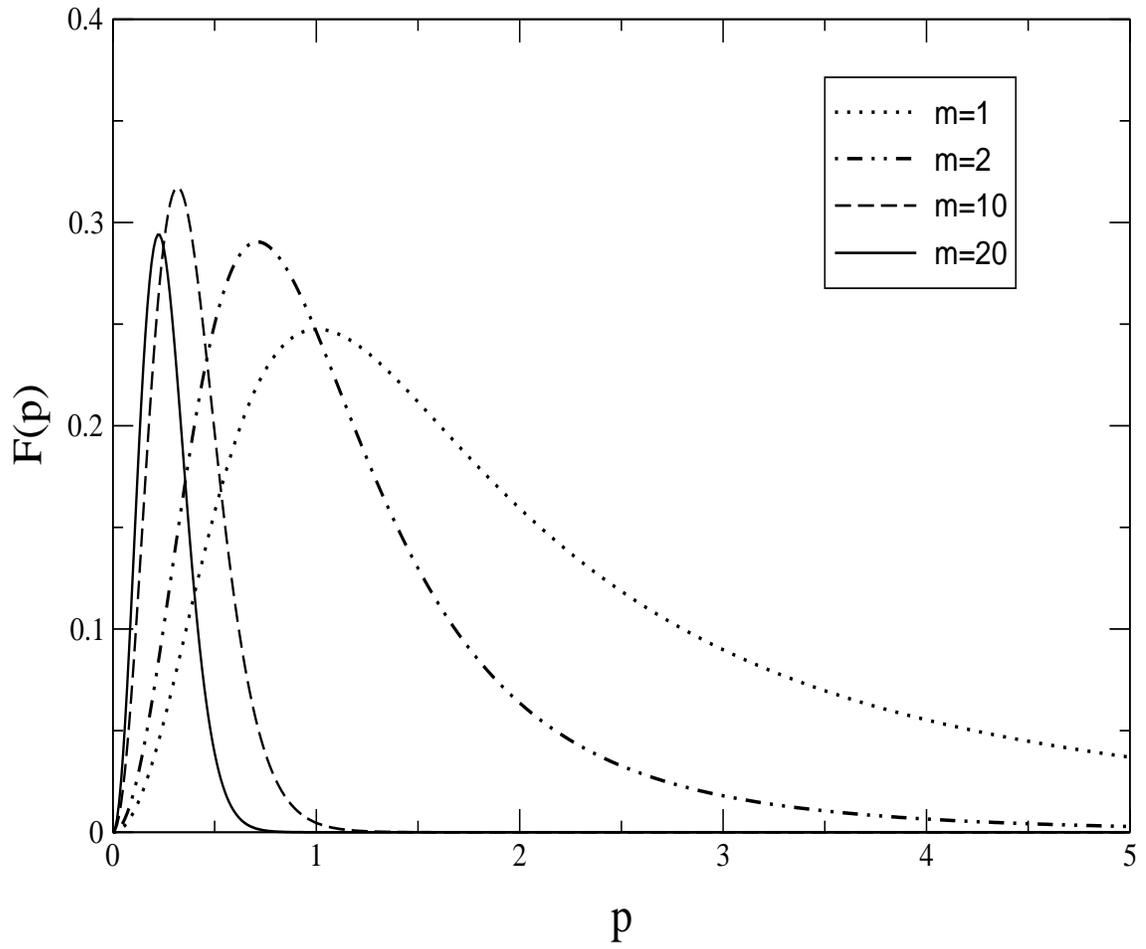}
\caption{
The function $F(p)$ (see text), plotted vs. $p$, for
fixed $k=1$, $V''=0.01$ and $m=1,2,10,20$.
}
\end{figure}

\end{document}